\documentclass[conference]{IEEEtran}
\IEEEoverridecommandlockouts
\usepackage{cite}
\usepackage{amsmath,amssymb,amsfonts}
\usepackage{algorithmic}
\usepackage{graphicx}
\usepackage{textcomp}
\usepackage{xcolor}
\usepackage{rotating}
\usepackage[export]{adjustbox}
\def\BibTeX{{\rm B\kern-.05em{\sc i\kern-.025em b}\kern-.08em
    T\kern-.1667em\lower.7ex\hbox{E}\kern-.125emX}}
\begin{document}

\title{Decoding the Popularity of TV Series: A Network Analysis Perspective\\
}

\author{\IEEEauthorblockN{Melody Yu}
\IEEEauthorblockA{\textit{Sage Hill School} \\ 
Newport Coast, California, USA \\}
}

\maketitle

\begin{abstract}
Viewer ratings are a key factor in shaping the success and continued viability of television shows. However, the factors contributing to high or low ratings are complex. This paper investigates whether character interactions in TV episodes, as captured through character network analysis, correlate with viewer ratings. Character networks are graphs created from the plot of a TV show that represent the interactions of characters in scenes. We constructed character networks from episode plots of three popular TV series and extracted graph metrics such as density and centrality measures from these networks. We explored the hypothesis that interactions between the characters in a TV episode, as quantified by the metrics of the episode character networks, are related to the overall reception of the episode, as indicated by the TV episode reviews. We undertook an analysis to examine the relationship between metrics of character networks and television show reviews. The results showed significant, though diverse, connections between specific network metrics and the ratings of episodes in different TV series. For example, in Game of Thrones, a negative correlation was observed between the count of active characters and the episode reviews, suggesting that episodes with too many characters might be less appealing to the audience. Conversely, Breaking Bad displayed a positive correlation between network transitivity and episode reviews, indicating that episodes with closely-knit character groups are generally more favored by viewers.  Our study contributes to understanding how character dynamics influence TV reviews, providing valuable insights for TV producers and writers in developing their shows. We demonstrated that network analysis can quantify character interactions. This enables producers to adjust character dynamics in future episodes to better engage audiences.

\end{abstract}


\section{Introduction}

People only have so much time. When we come home from work or school, we have our priorities; we do chores, send emails, or catch up on our favorite TV shows. For producers and TV channels, this time is imperative. The number of viewers on a show can vastly differ based on the time a show airs, and there are only so many slots available; ideal “prime times” like Tuesday at 9 pm can only be occupied by so many shows. For the 122.4 million households that watch TV, this poses an interesting question. What makes a show good; more so, what makes a show continue to be good? Unlike movies, TV shows consistently need to produce episode after episode, season after season. Sometimes viewership drops and your favorite 9 o’clock show is canceled and taken off-air or disappears from your favorite streaming service. But why? For the executives producing a show, the answer is simple: ratings. High ratings mean more seasons and low ratings mean the show ends. But there’s more to it than that. Quality fluctuates, and many shows are notorious for their lackluster attempts at good content in later seasons. What causes a great show, loved and revered in seasons one through three, to suddenly become an insult to its fans, its ratings plummeting in the next four seasons? To answer this question, we need to look at what makes a show good in the first place. Ratings are determined by viewer satisfaction, so what did I see in a show that made me want to invest my time in the first place? Is that related to why I don’t enjoy it now? 

The answer could be something that the viewers themselves might not even know. In this paper, we attempt to understand the psychology behind these ratings through character network analysis and study the relation between the character network metrics and TV viewership. We use different metrics of network analysis to quantitatively answer the question: do character interactions affect ratings in TV series?

From the interactions between characters in a TV episode, we construct a specific graph called character networks. Character networks are graphical representations of TV plots with characters as nodes and their interactions as connecting edges.  Upon establishing these networks, we apply network analysis techniques to quantify properties and patterns in the character interactions. By constructing and deriving metrics from these character networks for a show's episodes over time, we obtain quantitative insights into character dynamics. Our investigation then focuses on the correlation between these network metrics and episode reviews, testing the hypothesis that the nature of character interactions, as measured by network metrics, is related to TV episode reviews. Additionally, by correlating character network patterns with shifts in viewership throughout a show's duration, we aim to uncover potential explanations for changes in viewership.

\section{Related Work}

The study by Alberich et al. explored the network structure of the Marvel Universe, treating Marvel Comics characters as nodes and their co-appearances in comics as links \cite{marvel}. This work illustrates that even fictional social networks can exhibit properties similar to real-world social networks, like scale-free and small-world characteristics, offering insights into network theory applied in unconventional contexts.

In their research, Agarwal et al. studied the application of social network analysis to literary texts, specifically focusing on 'Alice in Wonderland'. \cite{alice} They constructed and analyzed networks based on the social interactions and events within the text, offering novel insights into character roles and relationships. This approach underscores the utility of network analysis in understanding the dynamics of literary characters and their interactions within a narrative framework.

Weng et al. presented an innovative approach to semantic movie analysis by constructing social networks of movie roles \cite{movieanalysis}. They demonstrated how the interrelationships between roles in a movie form a miniature society, providing significant clues for understanding the film. Their methodology involves constructing a roles' social network, identifying leading roles, and detecting hidden communities within this network. These elements contribute to storyline detection, facilitating advanced movie analysis and flexible browsing. The effectiveness of this approach is demonstrated through various applications and its robustness to errors in role recognition.

The paper by Kim et al. presents an innovative approach to analyzing social networks in literary fictions \cite{extract}. They demonstrate a method for constructing social networks from text by simple lexical analysis without complex natural language processing tools. Their findings reveal that these literary social networks exhibit power law distributions, a characteristic common in complex systems, and reflect the intrinsic narrative structures semantically designed by authors.

Bost et al. introduced 'narrative smoothing', a novel method for analyzing TV series plots by creating dynamic conversational networks  \cite{narritive}. This approach addresses the challenge of varying narrative pacing in TV series, providing insights into character relationships and storyline development over time. For this research, They published their dataset for three popular TV shows on \cite{data}. I utilized this publicly available dataset as a foundation to construct and advance the research topics in my study.

\section{Network Analysis}

Our research dataset is composed of character networks created by Bost. et. al. from three popular TV series: Breaking Bad (seasons 1-2), Game of Thrones (seasons 1-5), and House of Cards (seasons 1-2) \cite{data}. The source data comprises a sequence of segment graphs, each depicting character interactions across ten scenes. A node is defined as any character that speaks within this ten-scene window. The edges are thus defined as any conversational interaction between these nodes, where an undirected edge $E_{i, j}$ is drawn between any two characters $N_i$ and $N_j$ who speak directly to each other. Every edge  $E_{i, j}$  additionally carries a weight  $W_{i, j}$  representing the total conversation time between $N_i$ and $N_j$, expressed in seconds. Since each episode consists of multiple scenes, it is represented through several segment graphs. For example, in Game of Thrones episode 1, there are 31 segment graphs created from 310 scenes. Each segment graph of this episode contains between 14 and 17 nodes and 15 to 18 edges. 
 
\begin{figure}[htbp]
\centering
\includegraphics[width = \linewidth]{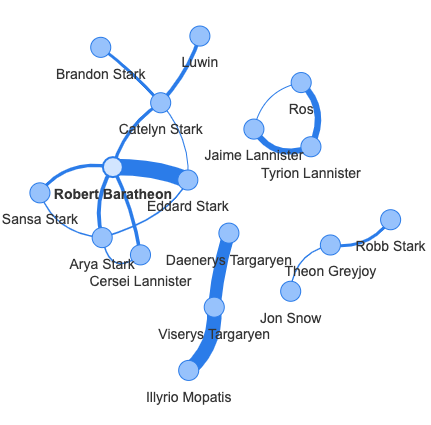}
\caption{Game of Throne Season 1, Episode 1, S120-S130 Character Network.}
\label{fig:graph}
\end{figure}

Figure~\ref{fig:graph} presents the character network episode graph from the TV show Game of Thrones season 1, episode 1. This character network is an undirected segment graph composed of four clusters. For example, in the first cluster, situated in the top left corner, describes a fully connected cluster of three characters. $N_1$ (Jaime Lannister) and $N_2$ (Tyrion Lannister) speak to each other, creating an undirected edge. $N_2$ is also directly connected to $N_3$ (Ros). The edges $E_{1, 2}$  and $E_{2, 3}$  have a similar weight, denoted by a similar thickness on the diagram. However, the edge between $N_1$ and $N_3$ has a smaller weight, shown as a thinner line. Weights are compared against all other clusters, with the average weight in some clusters shown as higher than in other clusters. For example, the average weight between the cluster containing $N_1$, $N_2$, and $N_3$ is higher than the cluster in the lower right corner of the diagram. This cluster includes $N_4$ (Jon Snow) speaking to $N_5$ (Theon Greyjoy). The edge $E_{4, 5}$ is similar in weight to $E_{2, 3}$, showing that the pairs of characters spoke for similar periods of time during these ten scenes. $N_5$ additionally spoke to $N_6$ (Robb Stark), connecting three nodes in a chain. $N_6$ (Robb Stark) and $N_4$ (Jon Snow) do not have a direct edge connecting them, as they did not directly speak to each other. 

To assess the popularity of TV shows, we incorporated IMDb's review scores, an online database offering information on films and television series. These reviews reflect the average episode rating, using a scale from 1 to 10, where higher scores indicate more favorable responses. IMDb aggregates these scores from individual reviews submitted by its registered users, with each user permitted one vote per episode. 

It is important to note that IMDb ratings for TV show episodes reflect the overall quality of the episode, not just the character interactions and plot. Other factors such as special effects, guest stars, and cinematography may also influence the score given by viewers. It is possible that these additional elements contribute to the overall enjoyment of the episode and are therefore included in the rating. 

\subsection{Graph Aggregation}\label{A1}
 
Each segment graph represents ten scenes from an episode, and every IMDb review reflects an entire episode. To analyze graph metrics for each episode, we combine all segment graphs from a single episode into one comprehensive weighted episode graph. Subsequently, we examine metrics of this aggregated graph and attempt to draw correlations between the IMDb ratings of episodes and various network indicators.

\begin{figure}[htbp]
\centering
\includegraphics[width = \linewidth]{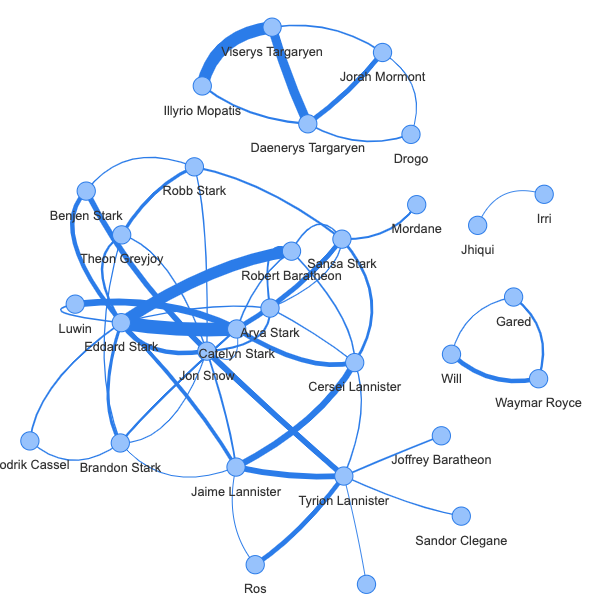}
\caption{Game of Throne Season 1, Episode 1, Aggregated Character Network.}
\label{fig:agg}
\end{figure}

The aggregation process works by iterating through edges in all segment graphs for a given episode. Edges connecting the same nodes will aggregate by summing their weights across all graphs. In essence, if an episode comprises 10 segment graphs and an edge appears in graphs 1, 4, 7, and 10, the weight $W_{i,j}$ of edge $E_{i,j}$ in our episode graph is calculated as the sum of the weights from these four segment graphs. This weight reflects the total conversation time in seconds between two characters in the episode. Figure~\ref{fig:agg} shows the aggregated character network for Game of Thrones season 1, episode 1. 

\[ W_{i,j} = W_{i,j}^1 + W_{i,j}^4 + W_{i,j}^7 + W_{i,j}^{10} \]

\subsection{Network Metrics}
We seek to establish a relationship between IMDb ratings and network indicators for a given TV series. Each temporal graph corresponds to ten scenes from an episode, whereas each IMDb review corresponds to one episode. Thus, to study the graph metrics for each episode, we first aggregate all segment graphs in one episode into one single weighted episode graph. Then, we study metrics on the aggregated episode graph.  

\subsubsection{Active Nodes}\label{A2}
Active nodes are the total number of nodes in a graph with at least one edge connecting them to another node. The number of active nodes represents the number of characters that speak during an episode. We use the number of active nodes in an episode graph to describe the complexity of the episode plot in this study. 

\subsubsection{Network Density}\label{A3}
A graph's density is the ratio between the number of actual edges that exist in a graph to the maximum number of edges that the graph can have. It indicates how many of a graph's potential connections are real connections. We define density as the following, where n represents the number of nodes and m represents the number of edges.  

\[D(G) = \frac{\textit{TotalEdges}}{\textit{MaxPossibleEdges}} = \frac{2 * m}{n(n - 1)}\]
 
The density of the graph increases as more characters converse with one another. Characters in sparse graphs are often still connected, but the conversations are between two characters and not necessarily talk with many other characters. We use network density to measure the level of character interactions in each TV episode.

\subsubsection{Node Strength}\label{A4}
A node’s strength is defined as the sum of the strengths, or weights, of the node’s edges. Since the edge weight in our graph indicates the time spent conversing between two characters, the strength of a character is the total amount of time spent conversing with all other characters. 

The maximum value of node strengths describes the total conversation time of the character who talked most with other characters during this episode. We use this metric to describe the exposure of the most active character in the episode. 

\subsubsection{Network Efficiency}\label{A17}
The efficiency of a graph refers to how efficiently nodes can reach each other.  It is also called communication efficiency. The efficiency of a pair of nodes in a graph is inversely proportional to their shortest distances: to find the efficiency of nodes $N_i$ and $N_j$,  we would define it as $\frac{1}{D_{i,j}}$  where $D_{i,j}$ is defined as the shortest distance between nodes $N_i$ and $N_j$.  The global efficiency of a graph is then defined as the average over the pairwise efficiencies. 

\[E(G) = \frac{1}{n(n - 1)} \frac{1}{\Sigma_{i\neq j}D_{i,j}}\]

In this study, the TV episode graphs are usually separated into separate subgraphs, often without any connections between these subgraphs. To more accurately represent these subgraphs, we use local efficiency, calculated from the average value of a subgraph’s global efficiency.

\subsubsection{Network Transitivity}\label{A6}
Network transitivity is the overall probability for the network to have interconnected adjacent nodes, revealing the existence of tightly connected communities. It refers to the extent to which the relation that relates two nodes in a network that are connected by an edge is transitive. The network transitivity is the fraction of all possible triangles present in $G$. It is computed as dividing total number of node triangles in the graphs by the total number of “triads” (two edges with a shared vertex)

\[T(G) = 3\frac{\textit{\# of triangles}}{\textit{\# of triads}}  \]

Network transitivity is widely used to examine the level of clustering in social network analysis.  In the content of character networks,  high transitivity means that the network contains communities or groups of characters that are closely interacting with each other. 

We use network transitivity to measure the extent of the complex relationship between characters, and their relation to the plot.

\subsubsection{Degree Centrality}\label{AA}
In a connected graph, centrality is used to measure the importance of various nodes in a graph. In this study, we use various centrality metrics to measure the conversational interaction patterns of different characters in each TV episode.

Degree centrality is the simplest measure of centrality. The degree of a node is defined as the number of edges or connections that a node has to other nodes. In the context of character networks in TV shows, the degree of a character represents the number of other characters that they interact with in a given episode. Figure 6 illustrates the time series of characters with the top 5 highest degrees in the character network from the first episode of Game of Thrones.

A node with a higher degree than the rest of the nodes may be considered a central or pivotal character, as they have a large number of interactions with other characters. As our episode graph is aggregated from multiple segment graphs, the large number of interactions can be spread over different scenes, or concentrated in a few scenes involving many characters. 
 
In this research, we will use the distribution of node degrees, including the maximum value and standard deviation, to describe the patterns of conversational interactions among the characters in a given episode.

\subsubsection{Closeness Centrality and Harmonic Centrality}\label{A7}
The closeness centrality of a node in a connected graph, which is determined as the reciprocal of the sum of the shortest distances between the node and all other nodes in the graph, is a measure of centrality in a network.  A node with a higher closeness centrality value is closer to all other nodes.  

\[C(u) = \frac{n - 1}{\Sigma_{v\neq u}D_{u,v}}\]

However, in our study, the TV episode graph is not necessarily a connected graph, as there can be some characters forming subgroups that do not interact with characters outside their subgroup. Therefore, we use harmonic centrality instead of closeness centrality to allow disconnected networks in our research. The harmonic centrality of a node u is defined as the sum of reciprocals of the shortest path distances from all other nodes to node u. Two nodes that are unreachable to each other has a distance of infinity.

\[H(u) =  \Sigma_{v\neq u}\frac{1}{D_{v,u}}\]

In our study, we use harmonic centralization distribution  (max value, standard deviation) to measure the extent to which the plot of the TV episode is concentrated on a single actor or a group of actors.

\subsubsection{Eigenvector Centrality}\label{A8}
While degree centrality measures the number of connections a node has, it does not take into account the influence of a node's neighbors on its own importance. To address this, we can use eigenvector centrality, which considers the importance of a node's neighbors in determining the node's overall importance. 

A node with a high eigenvector score is connected to many other nodes with high scores, indicating that it is connected to influential characters. In the context of character networks in TV shows, a character with a high eigenvector centrality is likely to be connected to characters who are also considered influential within the network. This can be useful in understanding the importance of a character within the context of the overall plot and character relationships.

\section{Results}
In this study, we calculated network metrics for three popular TV series: Game of Thrones, House of Cards, and Breaking Bad.

\subsection{Network Metrics }\label{A9}
We studied 22-26 episodes from the first three seasons of each show and calculated various network metrics for each episode. These metrics include density, efficiency, and transitivity, as well as node-level metrics such as degree, harmonic closeness centrality, and eigenvector closeness centrality.

For each metric, we also calculated the maximum and standard deviation values for each episode's character network. The results of these calculations are shown in Appendix Table~\ref{table:got},  ~\ref{table:hoc}, and  ~\ref{table:bb} , with each row representing a single episode and the first two columns indicating the episode number and review. The remaining columns display the various network metrics for that episode.

\subsection{Correlation between network metrics and episode reviews}
To determine whether there is a relationship between the network metrics we calculated and the IMDB reviews, we used correlation analysis. Specifically, we used the Spearman correlation method, as it is well-suited for analyzing ordinal data such as TV episode review scores. While the absolute value of a review score (e.g. 8.7) may not provide much insight on its own, it is generally accepted that an episode with a higher review score is of higher quality than one with a lower score. As such, the review scores can be considered ordinal, with higher values indicating better quality. By using the Spearman correlation method, we were able to examine the relationship between the network metrics and the episode reviews and determine if there is a correlation between them.

\begin{table}[htbp]
\caption{Spearman correlation of Game of Thrones episode reviews.}
\begin{center}
\begin{tabular}{|l|l|l|}
\hline
\textbf{Network Metrics} & $\textbf{\text{Correlation}}$ & $\textbf{\text{pValue}}$\\
\hline
\hline
Active Nodes & -0.49 & $\textbf{0.021}^*$ \\
\hline
Density & 0.254 & 0.253 \\
\hline
Efficiency & -0.128 & 0.571 \\
\hline
Transitivity & 0.126 & 0.576 \\
\hline
Max Strength & -0.142 & 0.528 \\
\hline
Std Strength & -0.062 & 0.783 \\
\hline
Max Degree & 0.053 & 0.813 \\
\hline
Std Degree & 0.089 & 0.695 \\
\hline
Max Harmonic & -0.095 & 0.674 \\
\hline
Std Harmonic & 0.007 & 0.976 \\
\hline
Max Eigen & 0.236 & 0.29 \\
\hline
Std Eigen & 0.561 & $\textbf{0.007}^{**}$ \\
\hline
\end{tabular}
\label{table:corr1}
\end{center}
\end{table}

Table~\ref{table:corr1} shows the results of the Spearman correlation analysis for the TV series Game of Thrones, examining the relationship between the network metrics and episode reviews. The analysis found two significant correlations: a negative correlation between the number of active nodes and episode review scores, and a positive correlation between the standard deviation of eigenvector centrality and episode review scores.

\begin{table}[htbp]
\caption{Spearman correlation of House of Cards episode reviews. 
}
\begin{center}
\begin{tabular}{|l|l|l|}
\hline
\textbf{Network Metrics} & \textbf{Correlation} & \textbf{pValue} \\
\hline
\hline
Active Nodes & -0.353 & 0.077 \\
\hline
Density & 0.418 & $\textbf{0.034}^{*}$\\
\hline
Efficiency & -0.387 & $\textbf{0.051}^{*}$ \\
\hline
Transitivity & 0.005 & 0.982 \\
\hline
Max Strength & -0.189 & 0.356 \\
\hline
Std Strength & 0.143 & 0.487 \\
\hline
Max Degree & -0.493 & $\textbf{0.010}^{**}$ \\
\hline
Std Degree & -0.408 & $\textbf{0.038}^{*}$  \\
\hline
Max Harmonic & -0.434 & $\textbf{0.027}^{*}$ \\
\hline
Std Harmonic & 0.360 & 0.071 \\
\hline
Max Eigen & 0.03 & 0.883 \\
\hline
Std Eigen & 0.499 & $\textbf{0.009}^{**}$ \\
\hline
\end{tabular}
\label{table:corr2}
\end{center}
\end{table}

\begin{figure}[htbp]
\centering
\includegraphics[width = \linewidth]{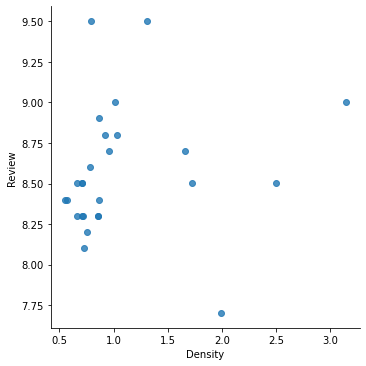}
\caption{Density vs Review for House of Cards}
\label{fig:degree}
\end{figure}

Table~\ref{table:corr2} shows the results of the Spearman correlation analysis for the TV series House of Cards, examining the relationship between the network metrics and episode reviews. The table indicates that there are six significant correlations between the two variables. However, we note that two of these correlations may be influenced by influential outliers, as visualized in Figure~\ref{fig:degree}. After accounting for these outliers, the analysis found four significant correlations: a negative correlation between network efficiency and episode review, a negative correlation between maximum node degree and episode review, a negative correlation between node degree standard deviation and episode review, and a negative correlation between maximum harmonic centrality and episode review. 

\begin{table}[htbp]
\caption{Spearman correlation of Breaking Bad episode reviews. 
}
\begin{center}
\begin{tabular}{|l|l|l|}
\hline
\textbf{Network Metrics} & \textbf{\text{Correlation}} & \textbf{\text{pValue}} \\
\hline
\hline
Active Nodes & -0.138	&  0.501   \\
\hline
Density & 0.162 &  0.429   \\
\hline
Efficiency & 0.300 & 0.136  \\
\hline
Transitivity & 0.448 & $\textbf{0.022}^{*}$  \\
\hline
Max Strength & 0.220 & 0.280 \\
\hline
Std Strength & 0.323 & 0.107 \\
\hline
Max Degree & 0.135 & 0.511 \\
\hline
Std Degree & 0.115 & 0.577 \\
\hline
Max Harmonic & -0.049 & 0.812 \\
\hline
Std Harmonic & 0.044 & 0.832 \\
\hline
Max Eigen & 0.061 & 0.765 \\
\hline
Std Eigen & 0.275 & 0.173 \\
\hline
\end{tabular}
\label{table:corr3}
\end{center}
\end{table}

Table~\ref{table:corr3} shows the Spearman correlation between episode network metrics and episode reviews for the TV series Breaking Bad. One significant correlation between the network metrics and episode reviews is a positive correlation between network transitivity and episode review.

\section*{Discussion}

The results of our research suggest that different character network structures may have different impacts on the quality of TV episodes. For example, the negative correlation between the number of active nodes and episode review scores in Game of Thrones may indicate that having too many characters in an episode can be overwhelming for viewers. On the other hand, the positive correlation between the standard deviation of eigenvector centrality and episode review scores suggests that focusing on a smaller number of main characters is preferred over a wider focus on many different characters in a single episode.

House of Cards had four correlations. The negative correlation between network efficiency and episode review implies that nodes are not in tight-knit groups - for example, character A might talk to character B who might talk to character C, but they might not talk in a group together. It also found a negative correlation between maximum harmonic centrality and episode review, implying again that characters have isolated conversations with one another rather than engaging in group discussions.

Breaking Bad had a single positive correlation between network transitivity and episode review. This implies that episodes with tightly connected groups of characters tend to be preferred by viewers.

It is important to note that a limitation of our study is that we only have the reviews for the episode as a whole, and not just for the character networks/dynamics. This means that many other factors such as cinematography, script writing, and plot, are all factored into a single review. Additionally, the placement and release of the episode are factors. For example, the finale of a series might be highly anticipated and receive a higher rating, or a guest star might appear in a particular episode and contribute to its score. Because the overall score of an episode represents all of these details together, we do not have the most accurate data to find a correlation between the character network and the metrics themselves. 

Our study is also limited to the types of metrics we tested. Though we tested many types, there are still many more network metrics that could potentially find greater correlations with the data.

\begin{table}[htbp]
\caption{Correlation between network metrics and reviews. 
}
\begin{center}
\begin{tabular}{|l|l|l|l|}
\hline
\textbf{Metrics} & \textbf{\text{Game of Thrones}}& \textbf{\text{House of Cards}} & \textbf{\text{Breaking Bad}} \\
\hline
\hline
Active Nodes & \textit{Negative} &  &  \\
\hline
Efficiency &  & \textit{Negative} &  \\
\hline
Max Degree &  & \textit{Negative} & \\
\hline
Std Degree &  & \textit{Negative} & \\
\hline
Transitivity &   &  & \textit{Positive} \\
\hline
Max Harmonic &  & \textit{Negative} & \\
\hline
Std Eigen & \textit{Positive} & & \\
\hline
\end{tabular}
\label{tab1e:result}
\end{center}
\end{table}

\section*{Conclusion}

This research aims to investigate the potential relationship between character interactions in an episode of a TV series and the review of that episode. We hypothesize that certain features of the TV series may attract viewers and lead to positive reviews if these features are present in the episodes. To capture some of these attractive features, we use character network analysis to analyze the interactions between characters in three well-known TV series. 

The correlation results of this study, shown in Table~\ref{tab1e:result}, indicate that there is a statistically significant correlation between character network metrics and TV show reviews. However, the specific network metrics that show significant correlation vary between the three series, with no common significant metric found across all three.

In summary, our research suggests that character networks can have an impact on TV show review scores. By analyzing the interactions between characters in TV series episodes using network metrics, we were able to identify statistically significant correlations between these metrics and review scores. These findings may be useful for TV producers and writers as they consider how to structure their shows and maintain audience engagement. While character networks are not the only factor that determines a show's success, they do play a role in audience enjoyment and should be carefully considered in the production of future seasons.

\onecolumn

\begin{sidewaystable}[htbp]
\caption{Network metrics for Game of Thrones episodes. 
}
\begin{center}
\small % Reduce font size
\begin{tabular}{|l|l|l|l|l|l|l|l|l|l|l|l|l|}
\hline
Episode & Review & Density & Efficiency & Transitivity & $Strength_{max}$ & $Strength_{std}$ & $Degree_{max}$ & $Degree_{std}$ & $Harmonic_{max}$ & $Harmonic_{std}$ & $Eigen_{max}$ & $Eigen_{std}$ \\
\hline\hline
1 & 9.001 & 0.690 & 0.612 & 0.403 & 91.350 & 21.330 & 10 & 2.510 & 13.833 & 4.248 & 0.627 & 0.157 \\\hline
1 & 9.001 & 0.690 & 0.612 & 0.403 & 91.350 & 21.330 & 10 & 2.510 & 13.833 & 4.248 & 0.627 & 0.157 \\\hline
2 & 8.701 & 1.083 & 0.598 & 0.371 & 108.371 & 27.805 & 10 & 2.293 & 14.667 & 3.516 & 0.651 & 0.169 \\\hline
3 & 8.601 & 0.559 & 0.627 & 0.391 & 123.114 & 26.931 & 14 & 2.662 & 17.500 & 3.073 & 0.600 & 0.136 \\\hline
4 & 8.701 & 0.510 & 0.470 & 0.332 & 153.735 & 32.168 & 12 & 2.413 & 15.500 & 3.117 & 0.693 & 0.140 \\\hline
5 & 9.001 & 0.817 & 0.388 & 0.304 & 241.514 & 47.983 & 12 & 2.228 & 19.283 & 3.789 & 0.612 & 0.146 \\\hline
6 & 9.101 & 0.808 & 0.378 & 0.293 & 146.025 & 37.887 & 13 & 2.103 & 18.067 & 4.035 & 0.702 & 0.149 \\\hline
7 & 9.201 & 0.832 & 0.437 & 0.340 & 156.197 & 34.976 & 10 & 1.636 & 10.500 & 2.189 & 0.606 & 0.152 \\\hline
8 & 9.001 & 0.312 & 0.565 & 0.418 & 58.668 & 16.305 & 8 & 1.885 & 11.667 & 2.303 & 0.702 & 0.135 \\\hline
9 & 9.601 & 0.554 & 0.581 & 0.455 & 233.417 & 39.914 & 8 & 1.850 & 11.500 & 2.271 & 0.704 & 0.137 \\\hline
10 & 9.501 & 0.374 & 0.346 & 0.308 & 80.360 & 19.776 & 9 & 1.756 & 14.000 & 3.544 & 0.707 & 0.140 \\\hline
11 & 8.701 & 0.539 & 0.364 & 0.345 & 142.963 & 31.816 & 8 & 1.654 & 14.500 & 3.680 & 0.649 & 0.129 \\\hline
12 & 8.501 & 0.524 & 0.344 & 0.293 & 189.477 & 38.741 & 8 & 1.578 & 17.200 & 4.666 & 0.705 & 0.129 \\\hline
13 & 8.801 & 0.424 & 0.399 & 0.336 & 214.796 & 32.493 & 9 & 1.592 & 16.667 & 4.384 & 0.698 & 0.127 \\\hline
14 & 8.701 & 0.534 & 0.481 & 0.369 & 118.127 & 28.484 & 9 & 1.862 & 11.000 & 2.387 & 0.628 & 0.136 \\\hline
15 & 8.701 & 0.585 & 0.409 & 0.331 & 145.672 & 29.029 & 8 & 1.753 & 9.500 & 1.824 & 0.688 & 0.133 \\\hline
16 & 9.001 & 1.035 & 0.512 & 0.396 & 171.112 & 40.708 & 7 & 1.482 & 8.500 & 1.534 & 0.666 & 0.157 \\\hline
17 & 8.901 & 0.686 & 0.291 & 0.330 & 107.020 & 30.333 & 6 & 1.355 & 8.500 & 1.713 & 0.694 & 0.148 \\\hline
18 & 8.701 & 0.752 & 0.301 & 0.304 & 223.797 & 41.606 & 6 & 1.296 & 8.000 & 1.869 & 0.698 & 0.143 \\\hline
19 & 9.701 & 0.727 & 0.337 & 0.358 & 104.391 & 27.663 & 9 & 2.206 & 12.783 & 3.962 & 0.652 & 0.168 \\\hline
20 & 9.401 & 0.670 & 0.249 & 0.323 & 116.497 & 28.533 & 5 & 1.140 & 7.250 & 1.616 & 0.696 & 0.145 \\\hline
21 & 8.701 & 0.465 & 0.352 & 0.291 & 189.694 & 34.213 & 8 & 1.487 & 8.500 & 1.787 & 0.705 & 0.132 \\\hline
22 & 8.501 & 0.459 & 0.401 & 0.367 & 109.471 & 31.511 & 7 & 1.340 & 10.417 & 2.058 & 0.548 & 0.125 \\\hline
\end{tabular}
\label{table:got}
\end{center}
\end{sidewaystable}

 \begin{sidewaystable}[htbp]
\caption{Network metrics for House of Cards episodes. 
}
\begin{center}
\small % Reduce font size
\begin{tabular}{|l|l|l|l|l|l|l|l|l|l|l|l|l|}
\hline
Episode & Review & Density & Efficiency & Transitivity & $Strength_{\text{max}}$ & $Strength_{\text{std}}$ & $Degree_{\text{max}}$ & $Degree_{\text{std}}$ & $Harmonic_{\text{max}}$ & $Harmonic_{\text{std}}$ & $Eigen_{\text{max}}$ & $Eigen_{\text{std}}$ \\
\hline\hline
1 & 8.601 & 0.785 & 0.235 & 0.096 & 217.296 & 42.013 & 17 & 3.111 & 23.000 & 2.684 & 0.691 & 0.156 \\\hline
2 & 8.501 & 0.662 & 0.345 & 0.153 & 184.978 & 33.518 & 19 & 3.220 & 25.000 & 3.065 & 0.682 & 0.142 \\\hline
3 & 8.301 & 0.720 & 0.346 & 0.109 & 241.030 & 42.640 & 22 & 3.700 & 25.833 & 4.243 & 0.669 & 0.144 \\\hline
4 & 8.201 & 0.756 & 0.330 & 0.170 & 191.937 & 36.833 & 17 & 3.042 & 22.333 & 3.866 & 0.677 & 0.147 \\\hline
5 & 8.401 & 0.548 & 0.504 & 0.246 & 193.091 & 34.697 & 20 & 3.587 & 26.667 & 3.591 & 0.674 & 0.145 \\\hline
6 & 8.501 & 0.712 & 0.370 & 0.174 & 246.106 & 46.610 & 19 & 3.324 & 25.333 & 3.036 & 0.673 & 0.150 \\\hline
7 & 8.101 & 0.726 & 0.343 & 0.173 & 197.469 & 39.098 & 18 & 3.233 & 25.500 & 3.061 & 0.606 & 0.141 \\\hline
8 & 7.701 & 1.995 & 0.465 & 0.421 & 162.628 & 44.609 & 8 & 1.765 & 8.500 & 2.157 & 0.706 & 0.213 \\\hline
9 & 8.501 & 1.726 & 0.487 & 0.301 & 208.798 & 47.180 & 12 & 3.320 & 17.000 & 2.465 & 0.658 & 0.164 \\\hline
10 & 8.701 & 1.655 & 0.494 & 0.256 & 181.880 & 44.724 & 14 & 3.054 & 17.000 & 2.170 & 0.563 & 0.156 \\\hline
11 & 9.001 & 3.141 & 0.310 & 0.173 & 225.517 & 55.327 & 8 & 1.983 & 11.500 & 1.604 & 0.679 & 0.185 \\\hline
12 & 8.501 & 2.497 & 0.365 & 0.210 & 264.034 & 63.197 & 9 & 2.285 & 13.833 & 1.893 & 0.706 & 0.205 \\\hline
13 & 8.801 & 0.917 & 0.251 & 0.152 & 168.177 & 33.982 & 12 & 2.683 & 18.667 & 2.331 & 0.692 & 0.163 \\\hline
14 & 9.501 & 1.309 & 0.302 & 0.240 & 183.422 & 40.448 & 11 & 2.496 & 16.833 & 2.254 & 0.669 & 0.167 \\\hline
15 & 8.301 & 0.707 & 0.290 & 0.145 & 197.214 & 36.916 & 20 & 3.542 & 25.500 & 3.046 & 0.646 & 0.144 \\\hline
16 & 8.401 & 0.568 & 0.231 & 0.221 & 198.008 & 33.332 & 15 & 2.617 & 24.167 & 3.208 & 0.692 & 0.141 \\\hline
17 & 9.001 & 1.015 & 0.401 & 0.266 & 200.016 & 50.566 & 11 & 2.397 & 17.917 & 4.145 & 0.676 & 0.167 \\\hline
18 & 8.301 & 0.856 & 0.430 & 0.218 & 244.365 & 47.852 & 16 & 3.059 & 23.167 & 4.651 & 0.661 & 0.156 \\\hline
19 & 8.301 & 0.858 & 0.414 & 0.188 & 246.606 & 45.328 & 14 & 2.662 & 22.667 & 3.072 & 0.670 & 0.152 \\\hline
20 & 8.401 & 0.865 & 0.310 & 0.210 & 216.038 & 40.224 & 14 & 2.873 & 22.167 & 2.982 & 0.686 & 0.152 \\\hline
21 & 8.501 & 0.708 & 0.271 & 0.195 & 207.177 & 38.964 & 14 & 2.684 & 22.500 & 4.098 & 0.680 & 0.153 \\\hline
22 & 8.901 & 0.863 & 0.226 & 0.219 & 164.527 & 44.760 & 9 & 2.273 & 19.500 & 2.697 & 0.551 & 0.154 \\\hline
23 & 8.301 & 0.660 & 0.339 & 0.251 & 234.251 & 44.020 & 14 & 2.774 & 20.417 & 4.202 & 0.645 & 0.144 \\\hline
24 & 8.701 & 0.962 & 0.232 & 0.170 & 234.172 & 47.400 & 12 & 2.420 & 19.917 & 3.765 & 0.683 & 0.154 \\\hline
25 & 8.801 & 1.034 & 0.272 & 0.194 & 155.699 & 31.591 & 10 & 2.013 & 18.000 & 3.441 & 0.589 & 0.150 \\\hline
26 & 9.501 & 0.791 & 0.270 & 0.172 & 206.005 & 39.814 & 14 & 2.713 & 23.000 & 2.862 & 0.688 & 0.156 \\\hline
\end{tabular}
\label{table:hoc}
\end{center}
\end{sidewaystable}

 \begin{sidewaystable}[htbp]
\caption{Network metrics for Breaking Bad episodes. 
}
\begin{center}
\small
\begin{tabular}{|l|l|l|l|l|l|l|l|l|l|l|l|l|}
\hline
Episode & Review & Density & Efficiency & Transitivity & $Strength_{\text{max}}$ & $Strength_{\text{std}}$ & $Degree_{\text{max}}$ & $Degree_{\text{std}}$ & $Harmonic_{\text{max}}$ & $Harmonic_{\text{std}}$ & $Eigen_{\text{max}}$ & $Eigen_{\text{std}}$ \\
\hline\hline
1 & 9.001 & 3.360 & 0.492 & 0.325 & 371.776 & 94.610 & 12 & 2.787 & 13.500 & 2.959 & 0.689 & 0.203 \\\hline
2 & 8.601 & 18.716 & 0.613 & 0.364 & 461.730 & 183.342 & 7 & 1.982 & 7.000 & 0.991 & 0.691 & 0.298 \\\hline
3 & 8.701 & 9.382 & 0.381 & 0.452 & 461.480 & 127.831 & 7 & 2.184 & 10.000 & 1.577 & 0.703 & 0.231 \\\hline
4 & 8.201 & 1.821 & 0.542 & 0.365 & 241.103 & 53.644 & 11 & 2.531 & 17.000 & 2.212 & 0.679 & 0.171 \\\hline
5 & 8.301 & 1.877 & 0.315 & 0.271 & 236.518 & 66.897 & 12 & 2.632 & 16.167 & 3.337 & 0.648 & 0.179 \\\hline
6 & 9.301 & 2.103 & 0.522 & 0.330 & 405.062 & 90.092 & 17 & 3.409 & 19.333 & 2.404 & 0.694 & 0.184 \\\hline
7 & 8.801 & 2.486 & 0.594 & 0.284 & 336.861 & 81.209 & 14 & 2.780 & 16.833 & 2.153 & 0.699 & 0.196 \\\hline
8 & 8.601 & 4.632 & 0.401 & 0.403 & 222.325 & 80.904 & 6 & 1.870 & 9.667 & 1.646 & 0.642 & 0.217 \\\hline
9 & 9.301 & 9.034 & 0.691 & 0.440 & 252.290 & 82.642 & 8 & 2.270 & 9.500 & 1.387 & 0.642 & 0.242 \\\hline
10 & 8.301 & 2.826 & 0.576 & 0.257 & 210.052 & 67.400 & 11 & 3.024 & 15.000 & 2.061 & 0.524 & 0.163 \\\hline
11 & 8.201 & 2.776 & 0.271 & 0.122 & 301.404 & 84.678 & 11 & 2.524 & 14.833 & 1.864 & 0.687 & 0.197 \\\hline
12 & 8.301 & 3.718 & 0.307 & 0.263 & 372.695 & 92.825 & 8 & 2.087 & 12.167 & 1.637 & 0.679 & 0.191 \\\hline
13 & 8.901 & 3.152 & 0.411 & 0.273 & 359.448 & 91.064 & 8 & 2.137 & 12.000 & 2.688 & 0.641 & 0.183 \\\hline
14 & 8.701 & 2.053 & 0.421 & 0.302 & 303.973 & 75.250 & 11 & 2.374 & 14.500 & 3.513 & 0.642 & 0.168 \\\hline
15 & 9.201 & 3.793 & 0.311 & 0.355 & 301.257 & 93.928 & 8 & 2.272 & 11.000 & 2.797 & 0.617 & 0.204 \\\hline
16 & 9.101 & 6.657 & 0.519 & 0.421 & 508.665 & 166.112 & 9 & 2.381 & 11.000 & 1.520 & 0.703 & 0.247 \\\hline
17 & 8.501 & 2.361 & 0.460 & 0.287 & 213.431 & 67.919 & 12 & 2.833 & 15.000 & 3.204 & 0.573 & 0.185 \\\hline
18 & 8.901 & 1.543 & 0.304 & 0.250 & 226.528 & 56.459 & 12 & 2.550 & 14.583 & 3.928 & 0.601 & 0.170 \\\hline
19 & 9.201 & 3.563 & 0.483 & 0.358 & 317.525 & 84.817 & 12 & 2.749 & 13.000 & 2.855 & 0.592 & 0.198 \\\hline
20 & 9.201 & 2.193 & 0.582 & 0.342 & 249.624 & 63.993 & 11 & 2.421 & 16.000 & 2.042 & 0.662 & 0.192 \\\hline
21 & 8.501 & 7.997 & 0.214 & 0.250 & 186.643 & 60.607 & 5 & 1.401 & 7.500 & 1.184 & 0.611 & 0.191 \\\hline
22 & 8.701 & 5.932 & 0.440 & 0.476 & 264.015 & 71.972 & 6 & 1.710 & 9.833 & 1.551 & 0.611 & 0.198 \\\hline
23 & 8.401 & 4.585 & 0.296 & 0.333 & 276.085 & 73.943 & 5 & 1.100 & 5.000 & 1.077 & 0.700 & 0.236 \\\hline
24 & 8.201 & 4.190 & 0.234 & 0.220 & 289.473 & 70.073 & 7 & 1.708 & 9.000 & 2.036 & 0.695 & 0.206 \\\hline
25 & 8.601 & 4.326 & 0.167 & 0.188 & 174.541 & 59.538 & 4 & 1.219 & 8.350 & 1.224 & 0.650 & 0.211 \\\hline
26 & 9.301 & 10.868 & 0.212 & 0.333 & 207.894 & 68.776 & 3 & 0.809 & 3.000 & 0.513 & 0.707 & 0.281 \\\hline
\end{tabular}
\label{table:bb}
\end{center}
\end{sidewaystable}

\end{document}